\documentclass[final]{IEEEtran}
\usepackage{amsthm,amssymb,graphicx,multirow,amsmath,color,amsfonts}
\usepackage[caption=false,font=footnotesize]{subfig}
\usepackage[update,prepend]{epstopdf}
\usepackage[noadjust]{cite}
\usepackage[latin1]{inputenc}
\usepackage{tikz}
\usepackage{bbm} 
\usepackage{pdfpages}
\usepackage{multirow}
\usepackage{comment}
\usepackage[ruled,vlined]{algorithm2e}
\usepackage{url}
\makeatother

\allowdisplaybreaks
\usepackage{setspace}

\newtheorem{prop}{Proposition}

\theoremstyle{definition}
\newtheorem{remark}{Remark}
\newtheorem{Def}{Definition}
\newtheorem{assum}{Assumption}

\DeclareMathOperator{\E}{\mathbb{E}}

\begin{document}
	
\title{Waveform Selection for Radar Tracking in Target Channels With Memory via Universal Learning \vspace{-.3cm}
{\footnotesize \textsuperscript{}}
\thanks{$^*$C.E. Thornton and R.M. Buehrer are with Wireless@VT, Bradley Department of ECE, Virginia Tech, Blacksburg, VA, 24061. (\textit{Emails: $\{$thorntonc, buehrer$\}$@vt.edu}). $^\ddagger$A.F. Martone is with the U.S Army Research Laboratory, Adelphi, MD 20783. (\textit{Email: anthony.f.martone.civ@mail.mil}). The support of the U.S Army Research Office (ARO) is gratefully acknowledged.}}
\author{\IEEEauthorblockN{Charles E. Thornton$^{*}$, R. Michael Buehrer$^{*}$, and Anthony F. Martone$^{\ddagger}$} \vspace{-.7cm}}

\maketitle
\begin{abstract}
In tracking radar, the sensing environment often varies significantly over a track duration due to the target's trajectory and dynamic interference. Adapting the radar's waveform using partial information about the state of the scene has been shown to provide performance benefits in many practical scenarios. Moreover, radar measurements generally exhibit strong temporal correlation, allowing memory-based learning algorithms to effectively learn waveform selection strategies. This work examines a radar system which builds a compressed model of the radar-environment interface in the form of a \emph{context-tree}. The radar uses this context tree-based model to select waveforms in a signal-dependent target channel, which may respond adversarially to the radar's strategy. This approach is guaranteed to asymptotically converge to the average-cost optimal policy for any stationary target channel that can be represented as a Markov process of order $U < \infty$, where the constant $U$ is unknown to the radar. The proposed approach is tested in a simulation study, and is shown to provide tracking performance improvements over two state-of-the-art waveform selection schemes.
\end{abstract}

\begin{IEEEkeywords}
Radar waveform selection, universal prediction, source coding, reinforcement learning, waveform diversity.
\end{IEEEkeywords}
\section{Introduction}
While tracking a moving target, a radar sequentially obtains information about the surrounding physical environment to meet the demands of a particular sensing objective. It has been well-established that dynamically varying the radar's waveform to match the behavior of the environment and current objective can result in appreciable performance improvements for both target detection and tracking \cite{Bell1993,Kershaw1994,Kershaw1997}. In these schemes, waveforms are selected based on a performance criterion of interest, such as minimum expected mean-squared tracking error \cite{Kershaw1994,Kershaw1997,Niu2002} or maximum mutual information between the target and received signal \cite{Bell1993}.

Many of the proposed waveform selection or design schemes rely on strict assumptions regarding the two-way propagation channel from the radar to the target and back. For example, in Bell's seminal work \cite{Bell1993}, an information-theoretically optimal scheme of \emph{probabilistically matching} the transmitted waveform to maximize information gain from a stochastic target is proposed. However, this scheme is only practical when the target channel is modeled as a Gaussian random process and the statistics of the target impulse response are known. Additionally, several works have proposed waveform selection based on a waveform specific Cram\'er-Rao Bound \cite{Kershaw1994,Kershaw1997}. However, these approaches are only viable in high SNR scenarios, which are rarely encountered in practical radar deployments. 

To devise general waveform selection strategies which do not heavily rely on \emph{a priori} knowledge of target channel behavior, reinforcement learning (RL) approaches have been proposed \cite{Thornton2020,PLiu2020,Thornton2020b}. In RL-based schemes, a model of the radar's environment is learned over time using closed-loop feedback. However, a notable limitation of RL is that problems are traditionally formulated as Markov decision processes (MDPs) to preserve Bellman optimality guarantees. In general, the memory length of the target channel may not be matched to the memory length assumed in a MDP. Computational challenges also arise as large, sparse matrices or neural networks are used to store model parameters.

Ideas from universal data compression can be applied to efficiently generalize the MDP model to a higher-order process. The well-known source coding algorithm of Lempel and Ziv \cite{Ziv1978} is commonly used to find an asymptotically optimal representation of any finite-order Markov source. Lempel-Ziv inspired algorithms have similarly been extended to develop universal prediction \cite{Merhav1998} and active learning \cite{Farias2010} schemes. We argue that universal source coding techniques can be used efficiently represent the radar-environment interference for optimal waveform selection in a broad class of environments.

\emph{Contributions:} We pose the radar waveform selection process as a partially observable stochastic control problem in Markov target channels of any finite order. We assert that the state-transition model can be represented as a \emph{context-tree}, which is an efficient model for encoding a stationary source with arbitrary memory. We develop a Lempel-Ziv based waveform selection algorithm, which is long-term cost optimal for any finite-order Markov target channel and compare tracking performance to a simpler contextual bandit learning algorithm and random waveform selection.
\section{Problem Formulation}
\label{se:formulation}
Consider a stationary and monostatic radar system located at the origin. The radar surveys a scene which contains at most one moving target. The scene may contain additional scatterers and interference from outside sources. The space of possible target states is represented as a two-dimensional grid in the delay-Doppler domain, where the delay cells are indexed by $\tau = \{1,...,M\}$ and the Doppler cells are indexed by $\nu = \{1,...,N\}$. Let the space of hypotheses regarding the target be denoted by $\mathcal{H}$, where $|\mathcal{H}| = (MN)+1$, with the additional hypothesis corresponding to `no target present' in the scene. The target's state is denoted\footnote{We denote by $x_{m}^{n}$ the sequence $\{ x_{m},x_{m+1},...x_{n} \}$.} by $x_{k} = i$, where the index $\{ k \in \mathbb{N}: 1 \leq k \leq K \} = [K]$ corresponds to discrete time and $i \in \mathcal{H}$ is a hypothesis regarding the target's location in the delay-Doppler space. The target's state evolves according to a finite-memory stochastic process with stationary transition probabilities 
\begin{equation}
	\mathbb{P} ( x_{k+1} = i|x_{k-L}^{k}) \quad \forall \; \; i \in \mathcal{H}, \; \; x_{k-L}^{k} \in \mathcal{H}^{L}, 
\end{equation}
where $L \in [0, \infty]$ is the memory length of the motion process, and the distribution exists for all possible values of $x_{k-L}^{k}$. Both the target state transition probabilities and memory length are unknown to the radar \emph{a priori}. The radar's goal is to determine the target's state with minimal uncertainty.

In addition to the target's state, the scene is also characterized by the state of the \emph{target channel} \cite{Bell1993}, given by $c_{k} \in \mathcal{C}$, where $|\mathcal{C}| < \infty$. Similar to the target's state, the state of the target channel evolves according to a finite-memory adaptive process, given by 
\begin{multline}
	\mathbb{P}(c_{k+1} = i | c_{k-J}^{k}, w_{k-J}^{k}), \quad \forall \; \; i \in \mathcal{C}, \\ c_{k-J}^{k} \in \mathcal{C}^{J}, w_{k-J}^{k} \in \mathcal{W}^{J} 
\end{multline}
where $w_{k}$ is the transmitted waveform at time $k$, selected from a finite alphabet, $|\mathcal{W}| < \infty$. The fixed constant $J \in [0,\infty]$ is the memory length of the channel-state generating process. It is important to note the dependence of the previous $J$ transmitted waveforms on the evolution of the channel state. Although some real-world emitters will not respond to the radar's choice of waveform, this model is general enough to consider coexistence with a reactive interfering system, which may co-operate or compete with the radar for channel resources. The state of the scene can then be viewed as the composition $s_{k} = [x_{k}, c_{k}]$ in set $\mathcal{S}$, having cardinality $|\mathcal{S}| = |\mathcal{H}|\times |\mathcal{C}|$. The scene transition probabilities are thus 
\begin{multline}
	\mathbb{P}(s_{k+1} = i|s_{k-U}^{k},w_{k-U}^{k}), \quad \forall i \in \mathcal{S}, \\ s_{k-U}^{k} \in \mathcal{S}^{U}, w_{k-U}^{k} \in \mathcal{W}^{U} 
\end{multline}
where $U = \max \{J,L\}$ is the memory length of the state generating process.

Instead of observing the true state $s_{k}$, the radar instead receives a noisy\footnote{To maintain tractability, we quantize the measurement alphabet to match the cardinality of the state space.} measurement $y_{k}$ from set $\mathcal{Y}$, where $|\mathcal{Y}| = 2^{NM} \times |\mathcal{C}|$. The probability of observing $y_{k+1} = j$ is given by the measurement model
\begin{multline}
	\mathbb{P} \left( y_{k+1} = j|s^{k+1}_{1} = i, w_{1}^{k} = h \right) \\ \quad \forall j \in \mathcal{Y}, i \in \mathcal{S}, h \in \mathcal{W},
\end{multline}
which reflects uncertainty about scene's state due to estimation errors. Since the radar does not directly observe $s_{k}$, it must make decisions based on the information available up until time $k$, given by $\mathcal{F}_{k} = \{w_{1}^{k},y_{1}^{k}\}$, which is the $\sigma$-algebra generated by the sequence of observations and actions, often referred to as the \emph{information state}. We assume the radar can store the entire information state in memory to enable knowledge gain. Since $\mathcal{F}_{k}$ contains all relevant information gathered until decision step $k$, it can be used to select waveforms in place of the true state sequence $\{s_{1}^{k}\}$. The main difficulty with using the information state to select waveforms is that the dimension of $\mathcal{F}_{k}$ grows linearly with $k$. To keep the problem tractable, the radar can utilize a sufficient statistic for $\mathcal{F}_{k}$. In this work, we make the following assumption: 

\begin{assum}
	\label{as:unbiased}
	Conditioned on the current information state, most recent measurement $y_{k}$ is an unbiased estimate of the true state $s_{k}$, ie. $\E[y_{k+1}|\mathcal{F}_{k}] = \E[s_{k+1}|\mathcal{F}_{k}]$. Thus $y_{k}$ is a sufficient statistic for $s_{k}$.
\end{assum}

\begin{remark}
	Assumption \ref{as:unbiased} is reasonable when the radar has detected the target from an earlier scanning period and obtains a measurement at each time step. In this case, both the target and spectrum observations can viewed as unbiased estimates of their true values.
\end{remark}

To calculate the measurement probabilities, simplifying assumptions can be made regarding the probability of detection and false alarm in each cell, as in \cite{LaScala2005}. However, we will not make any particular assumptions about the target scene or waveform catalog, and take the view that these probabilities must be \emph{learned} through repeated experience.

The sequence of transmitted waveforms can be interpreted as a being selected according to a policy, or decision function.
\begin{Def}[Policy]
	A policy $\mu$ is a sequence of mappings $\{\mu_{k}\}$, from which at each time $k$, the map $\mu_{k}: \mathcal{Y}^{k} \times \mathcal{W}^{k-1} \mapsto \mathcal{W}$ determines the waveform transmitted at time $k$ given the information state $\mathcal{F}_{k}$.
\end{Def}

After each waveform decision, the radar receives a bounded cost $g(y_{k},w_{k},y_{k+1}) \in [-g_{max},g_{max}]$, which quantifies the effect of $w_{k}$ on the uncertainty about the target's position\footnote{Considerations related to cost function design are discussed in Section \ref{se:cost}.}. To evaluate the performance of policy $\mu$, we define the long-term average cost
\begin{equation}
	\lambda_{\mu} \triangleq \lim_{K \rightarrow \infty} \E_{\mu} [ (1/K) \textstyle \sum_{k=1}^{K}g(y_{k},w_{k},y_{k+1}) ].
\end{equation}

Since the underlying state space is finite, the above limit always exists \cite{Bertsekas2006}. We can then define the optimal average cost over stationary policies by
\begin{equation}
	\lambda^{*} \triangleq \inf_{\xi} \limsup_{K \rightarrow \infty} \; \E_{\xi} [ (1/K) \textstyle \sum_{k=1}^{K}g(y_{k},w_{k},y_{k+1}) ],
\end{equation}
where the infimum is taken over the set of all admissible policies. The radar can then aim to find a policy $\mu$ that attains $\lambda^{*}$, provided that an appropriate cost function, which allows the radar to localize the target in the delay-Doppler grid of interest, is selected.

\begin{prop}[Structure of the optimal policy]
	For a state transition model of the form $\mathbb{P}(s_{k+1} = i|s_{k-L}^{k}, w_{k-L}^{k})$, an optimal policy $\mu^{*}$, which achieves the optimal long-term average cost $\lambda^{*}$ will be a Markov process of order $L$, eg. of the form $\mu_{k}(w_{k}|s_{k-L}^{k},w_{k-L}^{k})$.
\end{prop}  

The above proposition follows directly from the analysis of finite-state machine communication channels, where it is well-known that the optimal source distribution must be \emph{probabilistically matched} to the memory length of the channel. The interested reader can refer to \cite{Yang2005}, \cite{Tatikonda2009} for details regarding source-channel matching in finite memory channels.

Several practical considerations now arise. First, it is reasonable to assume the radar wishes to locate the target quickly and may not have the ability to extensively explore over the waveform library such that the transition kernel can be estimated. Thus, the need to either limit the state-action space or devise an efficient algorithm is clear. Secondly, the radar will obtain noisy state measurements, and will have to learn the memory length as well as the measurement model and transition probabilities to find $\mu^{*}$. Finally, traditional dynamic programming techniques assume a known memory length, and the model here has generalized this assumption to an unknown-order Markov process. 
\section{Universal Learning Algorithm}
Our algorithm is based on the active Lempel-Ziv algorithm proposed by Farias \emph{et al.} in \cite{Farias2010} and enhanced using the context-tree weighting method of Williems \emph{et al.} \cite{Willems1995}. This approach builds on ideas first presented in the context of universal source coding, in which the true distribution of a stationary source is approximated from a prior class of distributions by using a sliding memory window. In \cite{Ziv1978}, the idea of building a variable-length dictionary in the form of a context tree is proposed, where each node in the tree corresponds to a phrase that has been seen by the algorithm so far.  

The idea of building such a context tree\footnote{Due to space constraints, we forgo some of the formal development of context-trees. The interested reader is encouraged to consult \cite{Willems1995} for a thorough development.}, using pairs of observations and actions as nodes, was proposed for reinforcement learning in \cite{Farias2010}. We apply this general Lempel-Ziv inspired framework, in tandem with the context-tree weighting method of Willems \emph{et al.} \cite{Willems1995}, which is used to improve the rate at which the transition probabilities are estimated. 

The algorithm is seen in Algorithm \ref{algo:arlz}. At a high level, the algorithm splits time into \emph{phrases} of variable length. If the phrase covers the interval $[ \tau_{c}, \; \tau_{c+\ell} ]$, then the associated sequence of measurements and waveforms which characterizes the phrase will be $(s_{\tau_{c}}^{\tau_{c+\ell}}, w_{\tau_{c}}^{\tau_{c+\ell-1}})$, which corresponds to a node of the context tree. For each pair $s_{\ell+1} \in \mathcal{S}$ and $w_{\ell} \in \mathcal{W}$, the algorithm maintains an estimate of the transition behavior $\mathbb{P}(s_{\ell+1}|s^{\ell},w^{\ell})$, which is the probability of observing a particular state of the radar scene $s_{\ell+1}$ at the next time step when waveform $w_{\ell}$ is selected, given the current context $(s^{\ell},w^{\ell-1})$. The transition probabilities are initialized to a uniform distribution over the measurement space and updated using the observed counts of particular contexts. Let $N(s^{\ell+1},w^{\ell})$ be the number of times the context $(s^{\ell},w^{\ell})$ has been visited before step $k$. Then the transition probability can be estimated using the Krischevsky-Trofimov (KT) estimator
\begin{Def}[KT Estimator]
\begin{equation}
	\label{eq:kt}
	\mathbb{P}(s_{\ell+1}|s^{\ell},w^{\ell}) = \frac{N(s^{\ell+1},w^{\ell})+1/2}{\sum_{s' \in \mathcal{S}} N((s^{\ell},s'),w^{\ell})+|\mathcal{S}|/2},
\end{equation}
\end{Def}
which can be computed online using the observed frequency of each context at the current tree level.

\begin{remark}
	The compression performance of the KT estimator (\ref{eq:kt}) relative to the best constant probability assignment over $\mathcal{S}$ is bounded by $\frac{|\mathcal{S}|}{2} \log{K} + O(1)$, where $K$ is the total number of observations \cite{Merhav1998}.  
\end{remark}

While the KT estimator asymptotically minimizes the worst-case average redundancy, and can thus be used to estimate the state transition probabilities, convergence may be slow in general. To improve the rate of convergence, we apply the following weighting strategy.

\begin{Def}[Context-Tree Weighting]
	\begin{equation}
		\label{eq:ctw}
	P_{w}^{s} \triangleq \begin{cases}\frac{1}{2} P_{e}+\frac{1}{2} P_{u}^{0 s} P_{u}^{1 s}... & \text { for } 0 \leq l(s)<D \\ P_{e}, & \text { for } l(s)=D\end{cases},
\end{equation}
where $D$ is the current depth of the context-tree, $P_{e}$ is the KT estimate of node $s$, and $P_{u}^{0 s} P_{u}^{1s}...$ are the probabilities of the \emph{children} of node $s$.
\end{Def}

In addition to the transition probabilities, the \emph{cost-to-go} function must be estimated for each context \cite{Bertsekas2006}. The estimated cost $\hat{J}(s^{\ell+1},w^{\ell})$ is initialized to zero and subsequently updated using the rule
\begin{multline}
	\label{eq:ctg}
	\hat{J}(s_{\tau_{b}}^{s}, w_{\tau_{b}}^{s-1}) \leftarrow \underset{w_{s}}{\min} \textstyle \sum_{s_{s+1}} \mathbb{P}(s_{s+1}|s_{\tau_{b}}^{s}, w_{\tau_{b}}^{s-1}) \times \\ [g(s_{s},w_{s},s_{s+1}) + \gamma J(s_{\tau_{b}}^{s}, w_{\tau_{b}}^{s-1})],
\end{multline}
where $\gamma$ is a weighting term for prior estimates called the \emph{discount factor} and the update is performed by traversing backwards over the outcomes which have been previously observed, and the transition probabilities are estimated using (\ref{eq:kt}). Each step, the action is selected with the intent of either exploiting the action which is known to be most effective or gathering information about under-explored actions. This behavior is controlled by the sequence of exploration probabilities $\{\gamma_{k}\}$.

The universal learning approach is effective due to its general structure. Algorithm \ref{algo:arlz} will asymptotically converge to the long-term cost optimal for any stationary Markov decision process of finite order with discount factor $\gamma \approx 1$. The common MDP, contextual bandit, and multi-armed bandit problems are all contained within this umbrella.

\begin{algorithm*}[t]
	\setlength{\textfloatsep}{0pt}
	\label{algo:arlz}
	\caption{Adaptive Radar Lempel-Ziv Algorithm}
	\SetAlgoLined
	\textbf{Input} discount factor $\gamma$, sequence of exploration rates $\{\epsilon_{k}\}$\\
	Set $b \leftarrow 1, \tau_{b} \leftarrow 1$, $N(\cdot) \leftarrow 0 \; \; \textit{(context counts)}, \hat{J}(\cdot) \leftarrow 0$, $\hat{P}(\cdot) \leftarrow 1 / \lvert \mathcal{Y} \rvert$  \\
	\For{\text{Each CPI}}{
		\vspace{0.07cm}
		Radar observes $y_{k} \in \mathcal{Y}$; \textit{(Most recent measurement)}\\
		\eIf{$N(y_{\tau_{b}}^{k},w_{\tau_{b}}^{k-1}) > 0$ (\textit{known context})}{ 
			With probability $\epsilon_{k}$, select a random waveform $w_{k} \in \mathcal{W}$; (\textit{Exploration})\\
			OR with probability $1-\epsilon_{k}$, select greedy waveform with respect to $\hat{J}$; (\textit{Exploitation})	
		}{
			Select $w_{k}$ uniformly from $\mathcal{W}$; \textit{(Exploration in an unknown context)}\\
			\For{$u = k:-1:\tau_{b}$ (\textit{Traverse backwards and perform updates}) }{ 
				Increment $N(y_{\tau_{b}}^{u}, w_{\tau_{b}}^{u}) \leftarrow N(y_{\tau_{b}}^{u}, w_{\tau_{b}}^{u}) + 1$; (\textit{Update node count or add node to tree}) \\
				
				For each $y_{u} \in \mathcal{Y}$ update $\hat{P}(y_{u} | y_{\tau_{b}}^{s-1}, w_{\tau_{b}}^{s-1})$ using (\ref{eq:kt}); \\
				
				If node is not a leaf node apply (\ref{eq:ctw}) \\
				
				Update cost-to-go $\hat{J}(y_{\tau_{b}}^{u}, w_{\tau_{b}}^{u-1})$ using (\ref{eq:ctg});
				
			}
			$b \leftarrow b+1$, $\tau_{b} \leftarrow \tau_{b}+1$; (\textit{Begin the next phrase})
		}
		Radar receives cost $g(y_{k},w_{k},y_{k+1})$;
	}
\end{algorithm*}

\section{Cost Function Design}
\label{se:cost}
Due to the large body of research on the statistical theory of radar detection and estimation, many rigorous performance measures can be utilized. Generally, the problem of performance feedback is approached from either a control-theoretic or information-theoretic perspective. The former involves direct feedback from the tracking system to improve \emph{system level} as opposed to measurement level performance, and optimizes quantities such as the mean square tracking error or size of the target validation gate in measurement space \cite{Kershaw1994}. The information-theoretic perspective generally aims to maximize mutual information between the target and received signal, as in \cite{Bell1993}. However, to obtain a closed form expression, simplifications are often necessary.

For example, consider the problem of minimizing squared tracking error. In most cases, it is not possible to evaluate the mean square error (MSE) matrix analytically \cite{Bell2015}. Thus, it is common to use the \emph{Bayesian Cram\`er-Rao lower bound} in place of the MSE matrix. For target tracking, this involves conditioning on the observed data and computing the \emph{predicted conditional Cram\`er-Rao lower bound} (PC-CRLB). The PC-CRLB consists of a prior term and a data term. Unfortunately, the data term is difficult to compute in general, and it is common to assume a Gaussian measurement model.

In the information theoretic viewpoint, the target's impulse response is assumed to be a random vector $\mathbf{g}(t)$. If the radar transmits waveform $x(t)$, the resulting scattered signal $\mathbf{z}(t)$ is a finite-energy random process given by the convolution of $\mathbf{g}(t)$ and $x(t)$. Thus, a reasonable goal is to find waveforms which maximize the mutual information $I(\mathbf{g}(t); \mathbf{y}(t))$, where $\mathbf{y}(t)$ is the sum of $\mathbf{z}(t)$ and an additive noise process. The conditional mutual information $I(\mathbf{g}(t);\mathbf{y}(t)|x(t))$ is then easily computed if $\mathbf{g}(t)$ is a Gaussian process and the additive noise is Gaussian and independent of the transmitted waveform and target. Under these restrictive assumptions, Bell \cite{Bell1993} develops an optimal waveform design algorithm, based on the information-theoretic idea of waterfilling. Unfortunately, the proposed approach requires prior knowledge of the variance of $\mathbf{g}(t)$.

Under both viewpoints, modeling assumptions are required for tractable analysis. We can instead consider similar approaches, where the distributions are \emph{learned} over time by considering the relationship between particular waveform/observation pairs and the associated cost. The first objective function utilized is will be referred to as the tracking objective and is defined as follows.

\begin{Def}
	The tracking objective function is given by 
	\begin{equation}
		g_{\texttt{track}} \triangleq (\mathbf{Z}_{k}-\mathbf{\hat{X}}_{k})^{2},
	\end{equation}
	where $\mathbf{Z}_{k}$ is the current unfiltered measurement vector containing a range and velocity estimate for the target at time step $k$, and $\mathbf{\hat{X}}_{k}$ is the most recent target state estimate given by the tracking filter.
\end{Def}

Additionally, we propose an information theoretic objective function which seeks to minimize the negative entropy in the delay-Doppler image. This objective is defined as follows.

\begin{Def}
	The negative entropy objective function is given by
	\begin{equation}
		\label{eq:negentropy}
		g_{\texttt{entr.}} \triangleq \textstyle \sum_{i=1}^{N} \sum_{j=1}^{M} p_{ij}(k) \log(p_{ij}(k)),
	\end{equation}
	where the probability mass function $p_{ij}(k)$ is the probability that the target is located at delay-Doppler coordinate $(i,j) \in \mathcal{H}$ given the entire sequence of measurements. In practice, the probability of a target being present can be established using approximations, as in \cite{LaScala2005}, or by setting a detection threshold, and calculating a normalized distance from the energy in each cell to the threshold to establish a probability of the target being present. In this work, we opt for the latter approach. If the energy in a particular cell is very far from the detection threshold, then there is little ambiguity. Thus, minimizing (\ref{eq:negentropy}) will reduce uncertainty about the target's position.
\end{Def}

\begin{figure*}
	\centering
	\subfloat[]{\includegraphics[scale=0.45]{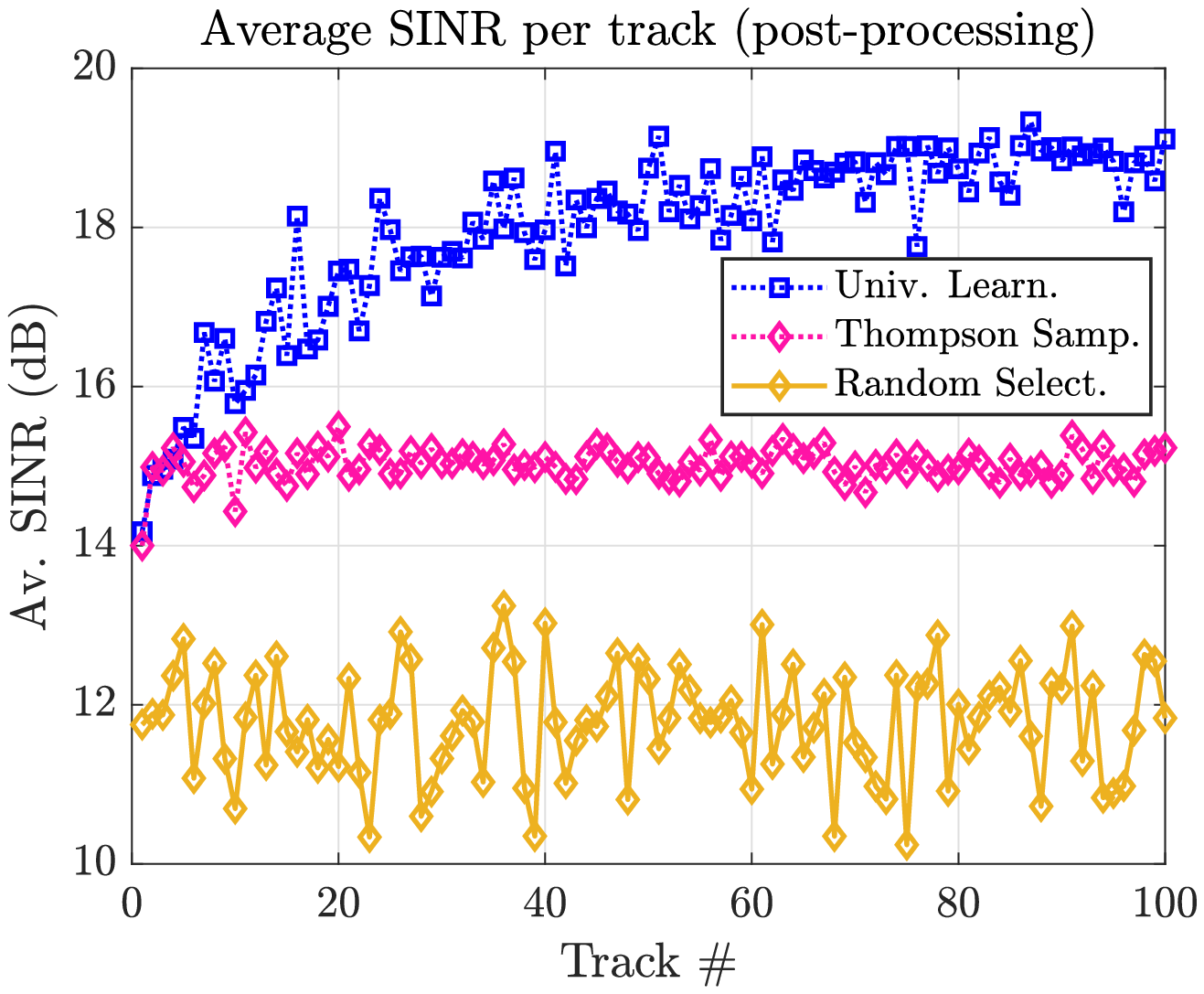}}
	\subfloat[]{\includegraphics[scale=0.45]{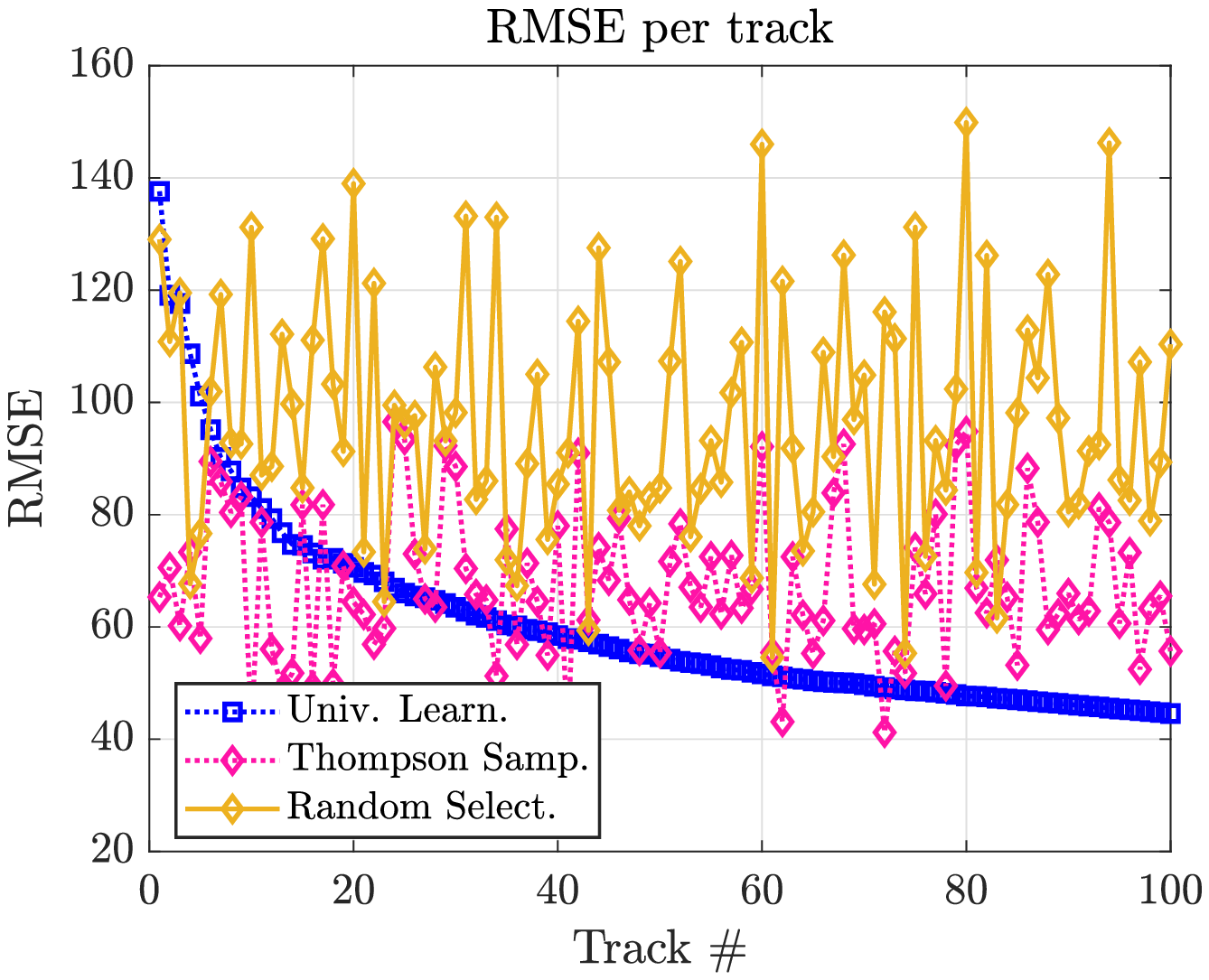}}\\
	\subfloat[]{\includegraphics[scale=0.45]{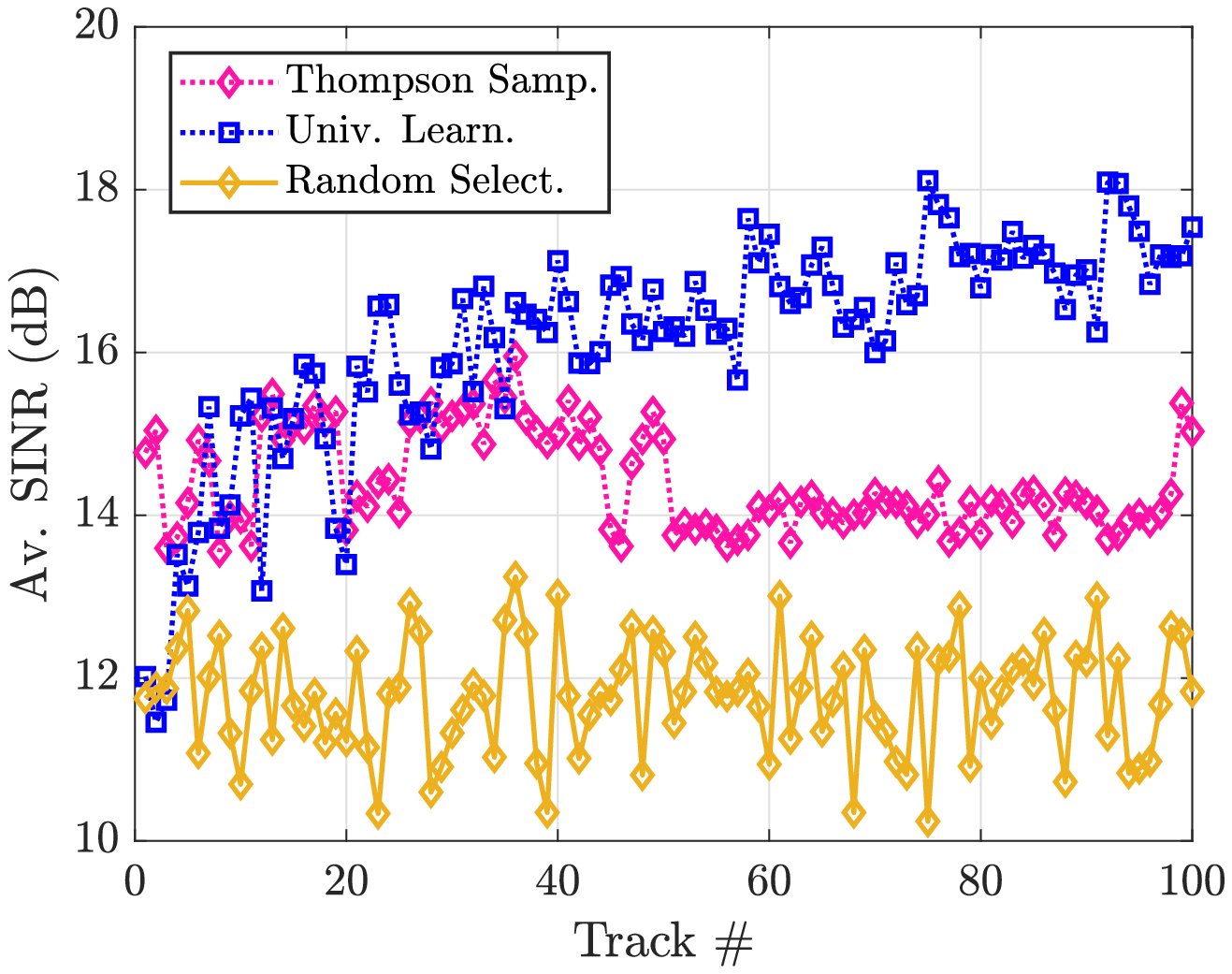}}
	\subfloat[]{\includegraphics[scale=0.45]{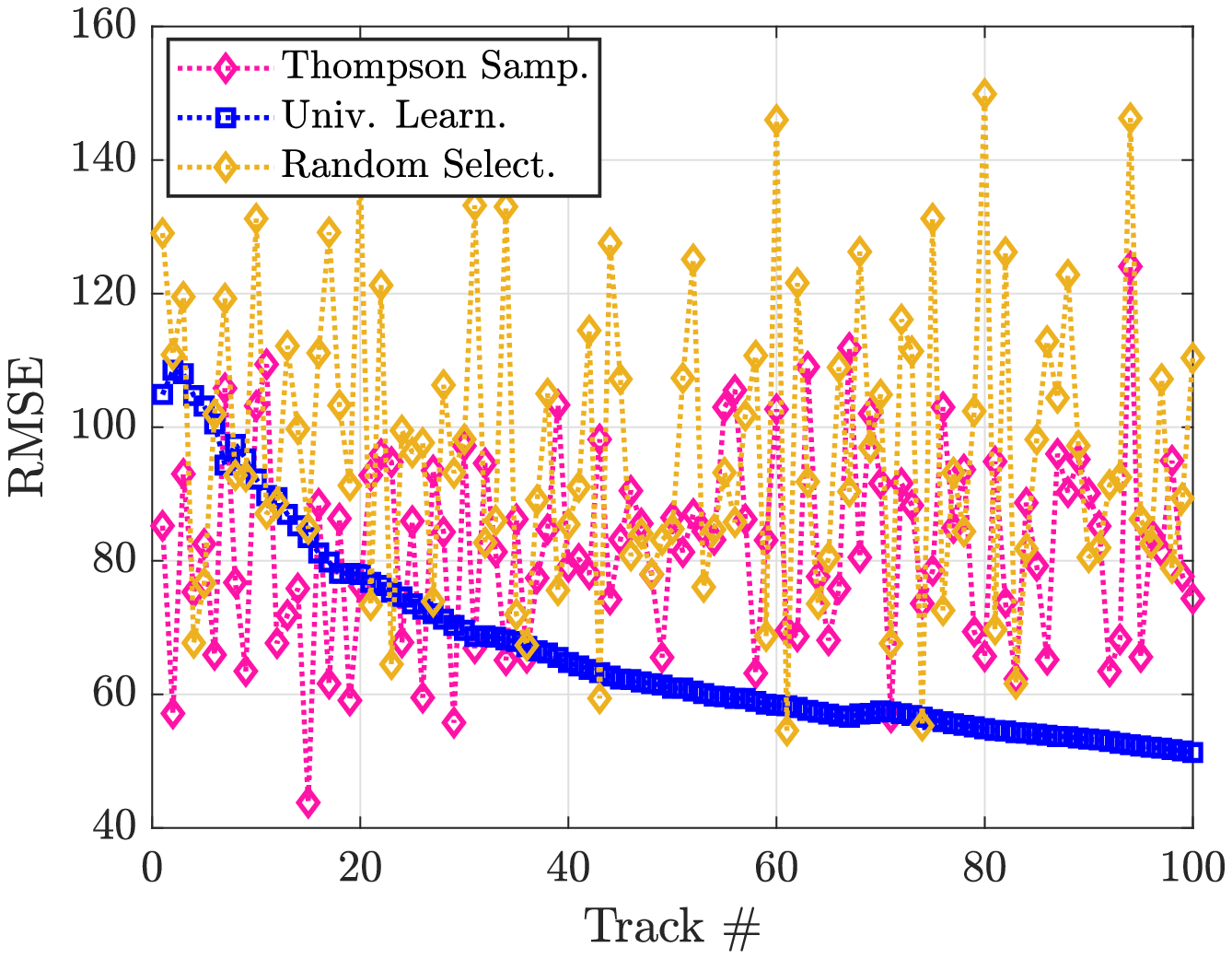}}
	\caption{\textsc{Average SINR and RMSE per track} under stochastic interference, which is drawn from a Markov chain of order 3. Each track consists of 200 CPIs in which the target is moving radially away from the radar at a constant velocity. In (a) and (b), the tracking objective function is used and in (c) and (d) the negative entropy objective is used.}
	\label{fig:stochControl}
	\vspace{-.3cm}
\end{figure*}

\section{Numerical Results}
\label{se:numerical}
In the following simulations, the radar is tracking a target which is moving radially away from the radar at a constant velocity. Each decision step corresponds to a Coherent Processing Interval (CPI) of $128$ pulses. The radar operates at a carrier frequency of $f_{c} = 2.5 \texttt{GHz}$ and employs a constant pulse repetition frequency of $0.496 \texttt{ms}$.

The hypothesis space consists of $1024$ delay cells and $512$ Doppler cells, which presents a total of $|\mathcal{H}| = (1024 \times 512) + 1$ possible target locations. The target channel state $c_{k}$ is represented by binary-valued vector of length $S$, which corresponds to the state of $S$ sub-channels that the radar may choose to transmit in. The radar's waveform catalog consists of LFM upsweep chirp waveforms as well as phase-coded waveforms using a Zadoff-Chu sequence of length $64$ in each of the possible sub-channels. Thus the cardinality of the waveform catalog is $2 \times S$. In the simulations, we compare the universal learning approach to Thompson Sampling (TS) based waveform selection, described in \cite{Thornton2020b}, as well as random waveform selection, which is a simple and commonly used technique in frequency-agile radar systems.

\begin{figure*}
	\centering
	\subfloat[]{\includegraphics[scale=0.45]{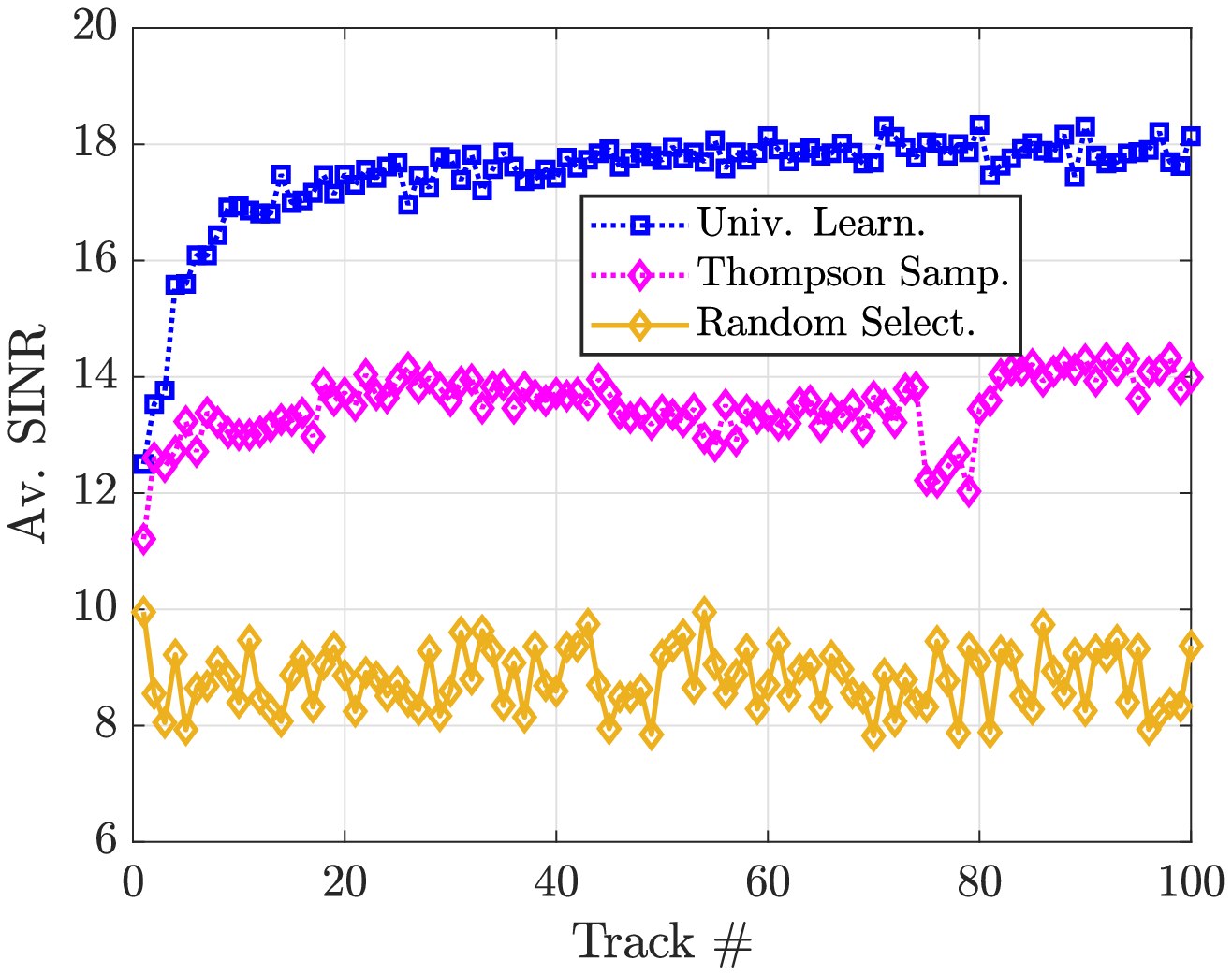}}
	\subfloat[]{\includegraphics[scale=0.45]{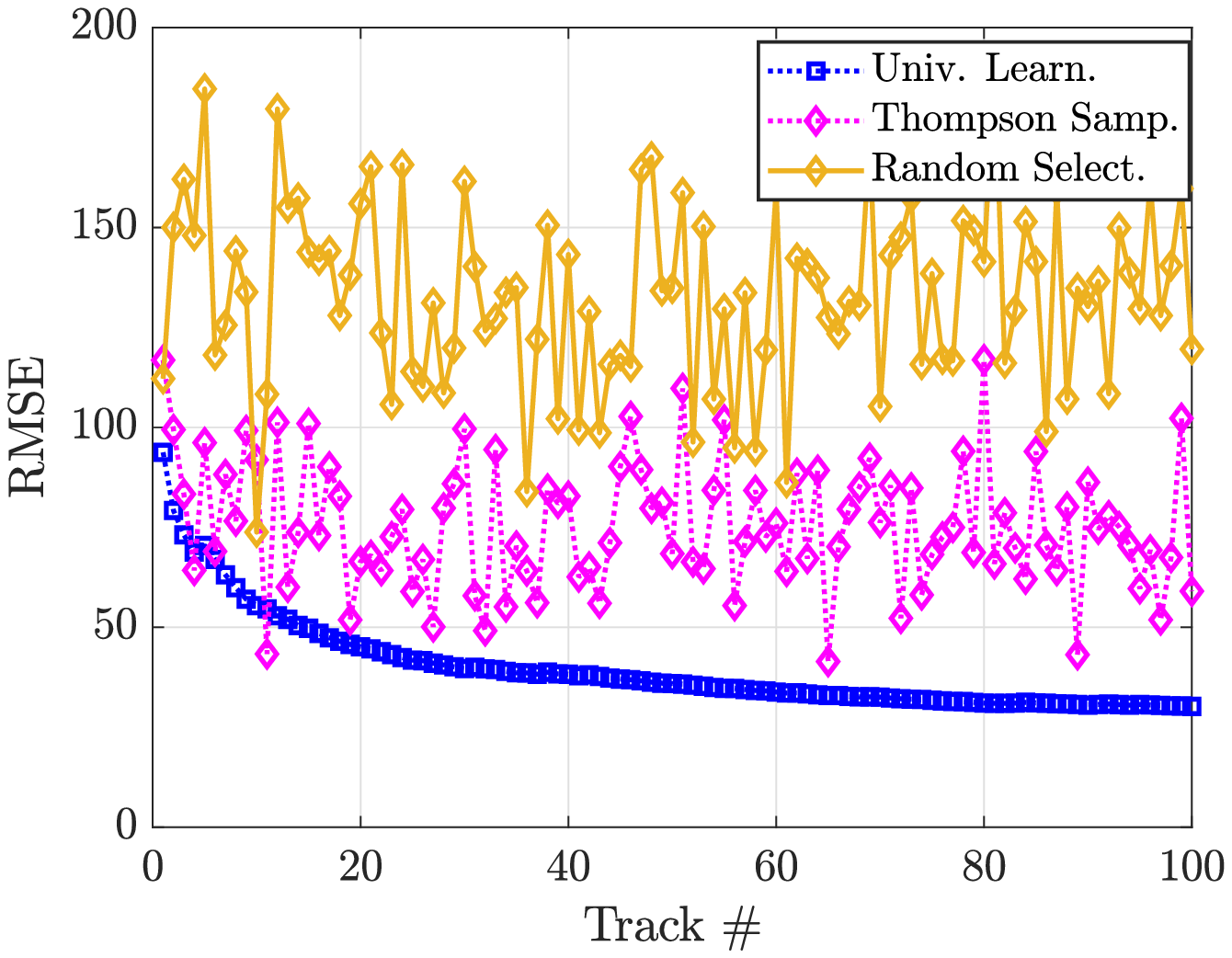}}\\
	\subfloat[]{\includegraphics[scale=0.45]{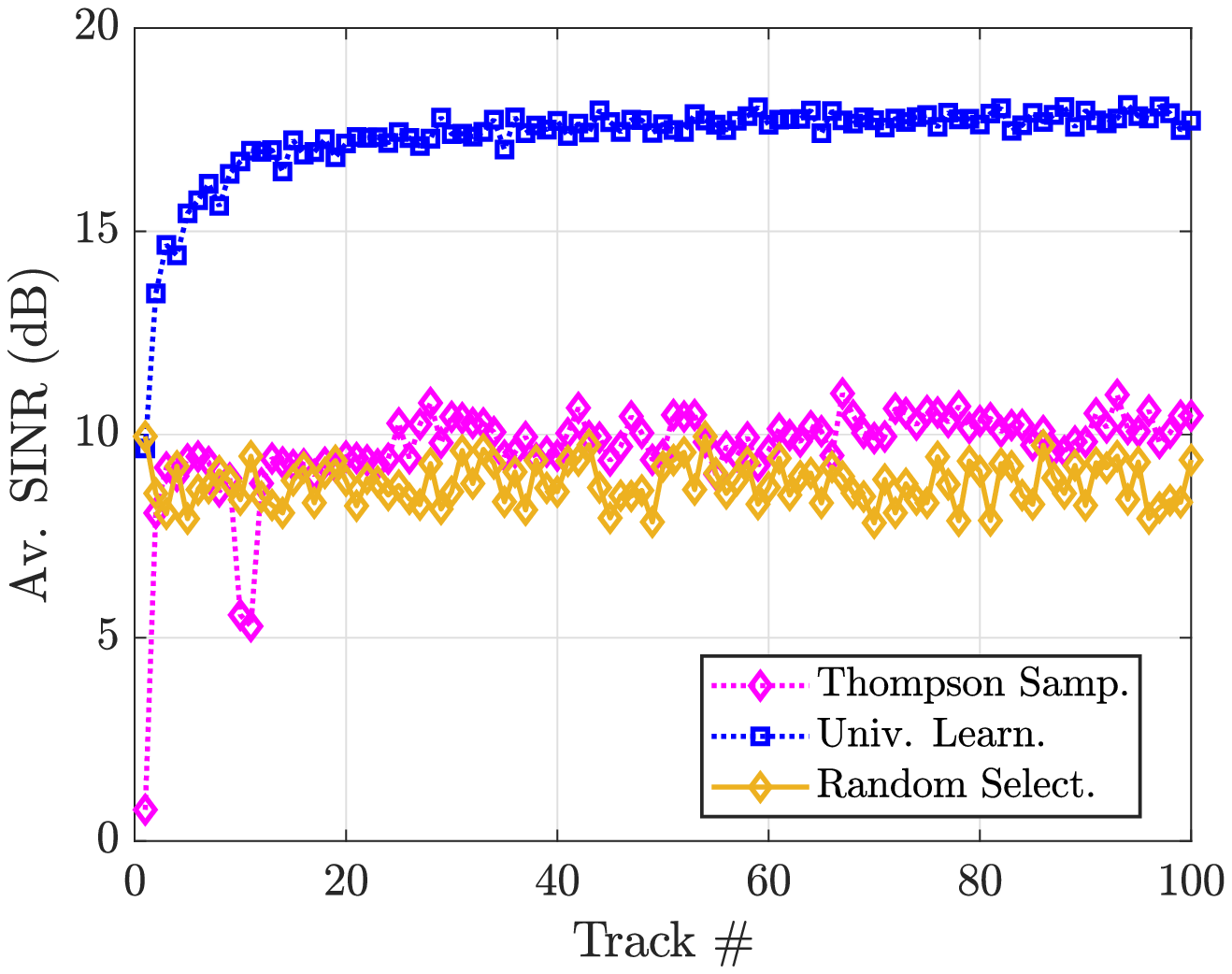}}
	\subfloat[]{\includegraphics[scale=0.45]{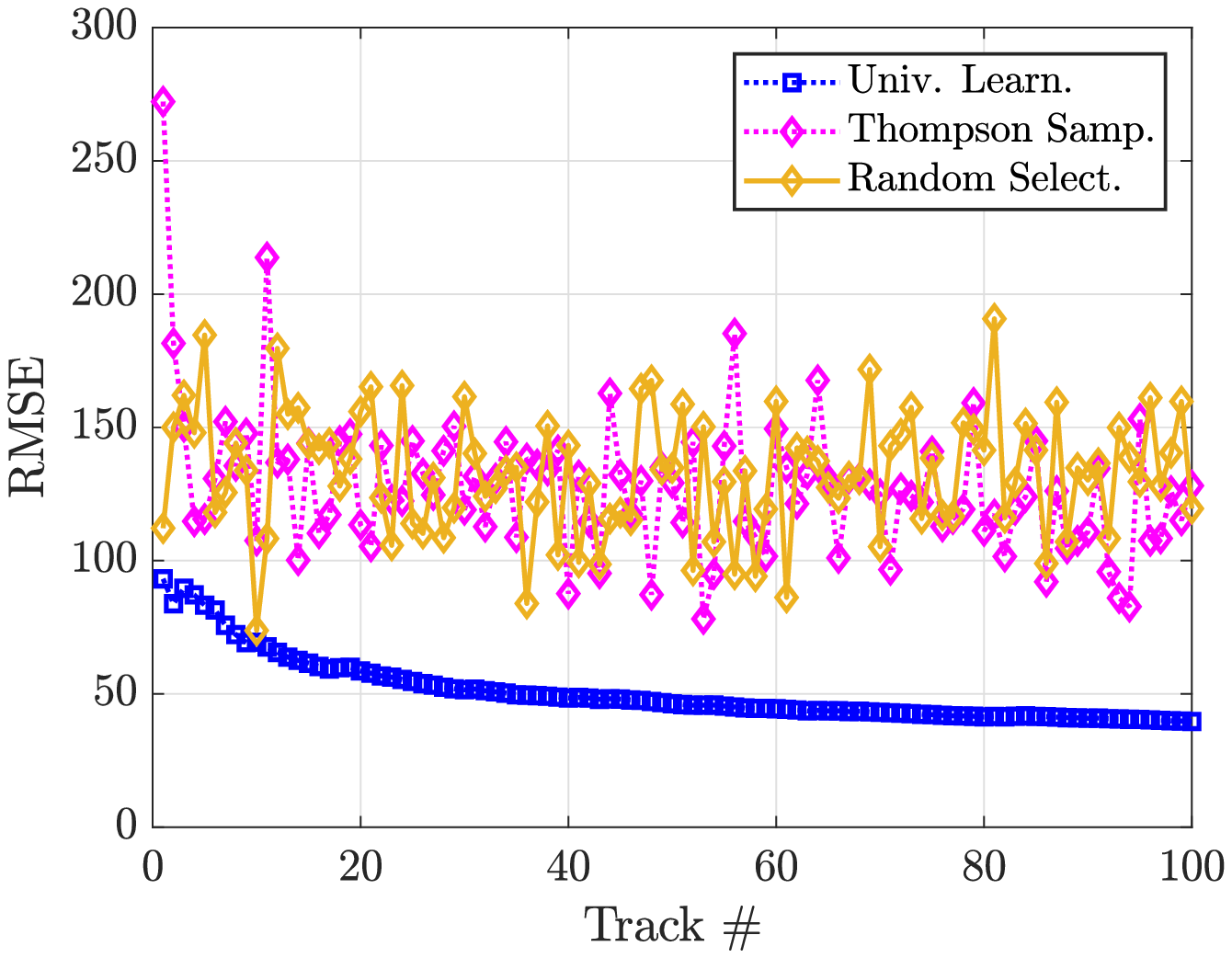}}
	\caption{\textsc{Average SINR and RMSE per track} under interference from an adaptive emitter, which is a Markov policy of order 2. Each track consists of 200 CPIs in which the target is moving radially away from the radar at a constant velocity. In (a) and (b) the tracking objective is used and in (c) and (d) the negative entropy objective is used.}
	\label{fig:jamIT}
	\vspace{-.3cm}
\end{figure*}

In Figure \ref{fig:stochControl}(a) and (b), the average measured SINR and tracking RMSE are observed over $100$ tracks, where each track consists of 200 CPIs and the radar uses the tracking objective function in the presence of stochastic interference, which is a Markov chain of length $3$. When the radar uses the universal learning approach, the radar requires a significant period for learning, but begins to approach the best possible tracking performance by the end of the $100$ tracks. In terms of $\texttt{SINR}$, performance is also favorable, as values above $13 \texttt{dB}$ correspond to very few missed detections. On the other hand, the (TS) approach converges to a stable solution by the second track, but performance remains stable for the remainder of the evaluation period, with $\texttt{SINR}$ stable at around $15 \texttt{dB}$ and a variable tracking error ranging from $40-95$. In terms of average $\texttt{SINR}$ both universal learning and TS provide a significant improvement over a random waveform selection policy, which is further confirmed by the improved RMSE performance also seen in Figure \ref{fig:stochControl}.

Additionally, we see the performance of each algorithm in Figure \ref{fig:stochControl}(c) and (d) when the negative entropy objective function is used. Results are fairly similar to the above case, but each algorithm performs slightly worse. A possible explanation for this behavior is that multiple waveforms which yield a similar expected reward for some contexts in terms of the negative entropy objective, which may differ in the case of the tracking objective. Once again, both the TS and Universal learning approaches provide a performance improvement over the random waveform selection policy.


In Figure \ref{fig:jamIT}(a) and (b), we examine a scenario where the radar's actions are tracked by an intelligent emitter. The radar is using the tracking objective function as the optimality criterion. While the Universal learning approach performs only marginally better than under the negative entropy objective, TS performs much better. Presumably, the algorithm is able to associate particular contexts with larger variations in tracking error more accurately than the relatively smaller variations in negative entropy. However, this could be dependent on the particular context representation used, which provides a degree of flexibility when implementing the TS approach.

Figure \ref{fig:jamIT}(c) and (d) show results from the adaptive emitter scenario when the radar is using the negative entropy objective function. In this scenario, if the radar transmits a particular waveform in the same frequency band for two consecutive CPIs, the emitter will transmit in that band during the next CPI. Otherwise, the emitter will remain in its current frequency band. The emitter is initialized to a random band. In this case, the universal learning algorithm learns much quicker than in the stochastic case. This is presumably because the emitter response to the radar is deterministic, given the radar's previous two waveforms and the emitter's current location. Additionally, the TS algorithm is not able to learn an effective policy, presumably due to the limitation in its context representation, which does not consider the radar's previously transmitted waveforms. 

\section{Conclusion}
\label{se:concl}
We examined the radar waveform selection process for target tracking under adaptive interference with an arbitrary memory length. The problem was formulated as a partially observable stochastic control problem with finite, but unknown memory and unknown state transition probabilities. To find an optimal policy, we proposed three schemes of varying complexity. We demonstrated each of the proposed schemes in simulation. We observed that the universal learning approach is a more effective utility minimizer than TS for the higher-order Markov environments tested here. Additionally, both algorithms resulted in a notable performance improvement over random waveform selection.

There are several avenues for continued investigation. A notable limitation of this formulation is that we have assumed an unbiased measurement model to maintain tractability. In future work, this model could be either generalized or estimated by an algorithm. Additionally, due to the high complexity of the universal learning algorithm, the size of the waveform catalog and state-space discretization was limited to maintain tractability. While this did not hinder performance for the cases examined here, it is possible for more realistic scenarios, additional diversity in the waveform catalog would be of major benefit. Finally, future work could focus on modeling more physical characteristics of the radar scene, such as additional scatterers, target trajectories, and RCS models.

\bibliographystyle{IEEEtran}
\bibliography{main}

\end{document}